\documentclass[twocolumn,prb,showpacs,preprintnumbers,superscriptaddress,amsmath,amssymb,citeautoscript]{revtex4-1}
\usepackage{graphicx}
\usepackage{dcolumn}
\usepackage{color}
\usepackage{bm}
\usepackage[hidelinks]{hyperref}
\hypersetup{
    colorlinks,
    citecolor=blue,
    filecolor=blue,
    linkcolor=blue,
    urlcolor=blue
}

\newcommand{\im}{\textrm{i}}

\DeclareMathOperator{\sgn}{sgn}

\DeclareMathOperator{\e}{e}
\DeclareMathOperator{\C}{\mathcal{C}}
\DeclareMathOperator{\T}{\mathcal{T}}
\DeclareMathOperator{\CL}{\mathcal{L}}
\DeclareMathOperator{\V}{\mathcal{V}}
\DeclareMathOperator{\M}{\mathcal{M}}

\begin{document}

\title{Visualizing Majorana bound states in 1D and 2D using the generalized Majorana polarization}

\author{N. Sedlmayr}
\email{nicholas.sedlmayr@cea.fr}
\affiliation{Institute de Physique Th\'eorique, CEA/Saclay,
Orme des Merisiers, 91190 Gif-sur-Yvette Cedex, France}
\author{C. Bena}
\affiliation{Institute de Physique Th\'eorique, CEA/Saclay,
Orme des Merisiers, 91190 Gif-sur-Yvette Cedex, France}
\affiliation{Laboratoire de Physique des Solides, UMR 8502, B\^at. 510, 91405 Orsay Cedex, France}

\date{\today}

\begin{abstract}
We study the solutions of generic Hamiltonians exhibiting particle-hole mixing. We show that there exists a universal quantity that can describe locally the Majorana nature of a given state. This pseudo-spin like two-component quantity is in fact a generalization of the Majorana polarization (MP) measure introduced in Ref.~\onlinecite{Sticlet2012}, which was applicable only for some models with specific spin and symmetry properties.
We apply this to an open two-dimensional Kitaev system, as well as to a one-dimensional topological wire. We show that the MP characterization is a necessary and sufficient criterion to test whether a state is a Majorana or not, and use it to numerically determine the topological phase diagram.
\end{abstract}


\maketitle

\section{Introduction}

Recently the formation of Majorana fermions has been a central research problem in condensed matter physics\cite{Kitaev2001,Nayak2008,Fu2008,Fu2009,Sato2009,Lutchyn2010,Oreg2010,Santos2010b,Alicea2011,Choy2011,Mourik2012,Deng2012,Das2012,Lee2014,Nadj-Perge2014}.
However, we believe that to address this problem a fundamental ingredient is lacking and the purpose of the present work is to introduce this missing piece and prove its importance. We thus introduce the generalized Majorana polarization (MP), which is a universal measure of the spatial dependence of the Majorana 
 of a given state, i.e. of the same-spin particle-hole overlap. This quantity can be thought of as the analog of the local density of states (LDOS) for regular electrons, except for the fact that a real quantity does not suffice to capture the Majorana character, and one needs to introduce a complex quantity which can be represented as a 
two-component vectorial quantity in the complex plane.

It turns out that this quantity can be obtained from the particle-hole (PH) operator expectation value.Such an operator $\C$, is anti-unitary, obeys $\C^2=1$, and anti-commutes with the Hamiltonian.\footnote{Such models are referred to as D, BDI and DIII in the usual classification scheme\cite{Schnyder2008}.}
The Majorana bound states (MBS) are self adjoint, i.e.~they are eigenstates of the PH operator with an eigenvalue of modulus 1. It is therefore natural to use the MP which stems from the PH operator to analyze the MBS in more detail.
Note that the MP vector discussed here is a generalization of the MP introduced in Ref.~\onlinecite{Sticlet2012} which was applicable only for a specific subset of models.\footnote{In particular it was applicable to the `chiral orthogonal' class, or BDI,\cite{Schnyder2008} class: the BDI models have, in addition to the particle hole symmetry, a chiral symmetry obeying $[H,\T]=0$ and $\T^2=1$, where $H$ is the Hamiltonian. In addition the former MP required specific spin densities.}

Having access to such a local measure can allow one to understand the evolution of these states through a phase transition, their dependence on specific particularities of the system such as size, disorder, inhomogeneities, etc., as well as how one can manipulate them. 
The particular patterns arising in the spatial distribution of the MP vector, i.e spatially aligned (`ferromagnetic'), vortex-like, localized-on-the-edges, etc, and its integral over given regions in space, allow one to assign a global topological character for any given state. 

In what follows we write down the generalized MP definition and apply it to a few examples, such as two-dimensional $p+\im p$-wave Kitaev arrays, and a topological one-dimensional wire in the presence of various inhomogeneities. 
As we will show, the spatial distribution of the MP vector, allows one to distinguish between states that exhibit a trivial or topological 
phase.
When the criterion of a zero energy for a given state cannot be strictly applied (e.g infinitesimally small but non-zero energies), having a universal local order parameter is a sufficient and versatile criterion for such a distinction. This allows the accurate determination of the topological phase diagram from numerical calculations.

Also we find analytically the phase diagram for quasi 1D Kitaev wires using an exact calculation of the topological invariant for these systems. The value of this topological invariant, and the corresponding phase diagrams were previously unknown. We compare the phase diagrams obtained using the two techniques and we note that the MP criterion works very accurately, even for not too large systems. Thus, the MP is of potential use for the determination of the topology of more complicated realistic models, for which the direct determination of the topology using the topological invariant is unfeasible.

Moreover, as we will show, this criterion will help us prove the existence of quasi-Majorana or precursor Majorana states, which are locally but not globally Majorana-like. Such states exhibit locally an almost perfect electron-hole superposition, thus a quasi-maximal MP, however the direction of the MP vector may vary spatially and thus one cannot isolate a well-defined region that would integrate to a fully localized Majorana state.

In Sec.~\ref{sec_mp} we present the definition of the generalized MP. In Sec.~\ref{sec_kit} we apply this definition to quasi-1D and finite-size 2D system described by the Kitaev model. In Sec.~\ref{sec_spin_def} we apply it to a 1D spinful system. In Sec.~\ref{oldmpdef} we compare the present definition of the MP with the original definition in Ref.~\onlinecite{Sticlet2012}. We conclude in Sec.~\ref{conc}. In Appendix \ref{appa} we present the analytical calculation of the topological invariant for the quasi-1D Kitaev chains, while in Appendix \ref{app_char} we present the relation between the MP and the chiral character presented in Ref.~\onlinecite{Sedlmayr2015a}.

\section{The generalized Majorana polarization}\label{sec_mp}
Naturally all finite energy eigenstates of the Hamiltonians under consideration satisfy $\langle\Psi|\C|\Psi\rangle=0$ and Majorana states satisfy $|\langle\gamma|\C|\gamma\rangle|=1$, where $\C$ is the PH operator. 
Additionally in a region $\mathcal{R}$ where such a Majorana state is localized it must satisfy
\begin{equation}\label{mp}
C=\frac{\left|\sum_{j\in \mathcal{R}}\langle\Psi|\C_j|\Psi\rangle\right|}{\sum_{j\in \mathcal{R}}\langle\Psi|{\hat r}_j|\Psi\rangle}=1\,,
\end{equation}
where ${\hat r}_j$ is the projection onto site $j$ and $\C_j\equiv \C {\hat r}_j$. For a system with two Majoranas localized each on a different edge of the system, $\mathcal{R}$ can simply be taken to be half the system.

One can therefore use the PH operator as a way of theoretically visualizing MBS; this operator thus plays the role of a universal MP generalizing the picture introduced in Ref.~\onlinecite{Sticlet2012} which was valid only for a subset of Hamiltonians (the `chiral orthogonal' class, or BDI in the usual classification scheme\cite{Schnyder2008}). The relationship between the old definition and the current one is presented in Section~\ref{oldmpdef}.

Note that, while presenting some similarities, the general MP is different from the chiral Majorana character introduced in \onlinecite{Sedlmayr2015a}; they happen to take a similar form only for the particular case of the BDI systems, see App.~\ref{app_char} for more information. As the expectation values for an anti-unitary operator are not invariant under a change of global phase, it is fundamentally impossible to use the MP operator to compare different states, as we can do using the chiral Majorana character introduced in Ref.~\onlinecite{Sedlmayr2015a}.

\section{Kitaev chains, ladders, and arrays}\label{sec_kit}
The PH operator for a spinless Kitaev model is $\C=\e^{\im\zeta}{\bm \tau}^x\hat K$, where $\hat K$ is the complex-conjugation operator, and $\zeta$ is an arbitrary phase. We use $\vec{\bm\tau}$ to denote the Pauli matrices in the PH subspace. A Majorana state $\gamma$ is an eigenstate of $\C$ with an eigenvalue of modulus 1. If we write a general eigenfunction as 
$\Psi_j^T=(u_j,v_j)$, then $\langle\Psi|\C_j|\Psi\rangle=2u_j v_j$,
and we can use this to analyze the behavior of a given state and its Majorana character. Note that in order to have a Majorana state localized in a region $\mathcal{R}$ , the wave function must satisfy the condition $u_j=v_j e^{i \phi_j}$, with $\phi_j=\phi$ a uniform phase inside $\mathcal{R}$. This phase is arbitrary and can be chosen conveniently.

In the past, the topological character of a variety of quasi-1D and 2D systems has been studied.\cite{Potter2010,Potter2011,Stanescu2011,Law2011,Stanescu2013,Wang2014,Seroussi2014,Poyhonen2014,Wakatsuki2014,Thakurathi2014,Sedlmayr2015} Here we focus on a spinless 2D square lattice with nearest neighbor hopping $t$ and $p+\im p$ superconductivity of strength $\Delta$, described by
\begin{eqnarray}\label{kit_ham}
H&=&\sum_{j}\Psi^\dagger_{j}\mu{\bm\tau}^z\Psi_{j}\\\nonumber&&+\sum_{\langle i,j\rangle}\Psi^\dagger_{i}\left[
\Delta\left([\vec\delta_{ij}]^x\im{\bm\tau}^y+[\vec\delta_{ij}]^y\im{\bm\tau}^x\right)-t{\bm\tau}^z\right]\Psi_{j}\,.
\end{eqnarray}
where $\Psi^\dagger_{j}=\{c^\dagger_{j},c_{j}\}$ with $c_{ j}^{(\dagger)}$ annihilating (creating) a spinless particle at site $j$. Here $\vec\delta$ is the nearest neighbor vector. We set $t=1$ and $\hbar=1$ throughout.

\begin{figure}
\includegraphics*[width=0.45\columnwidth]{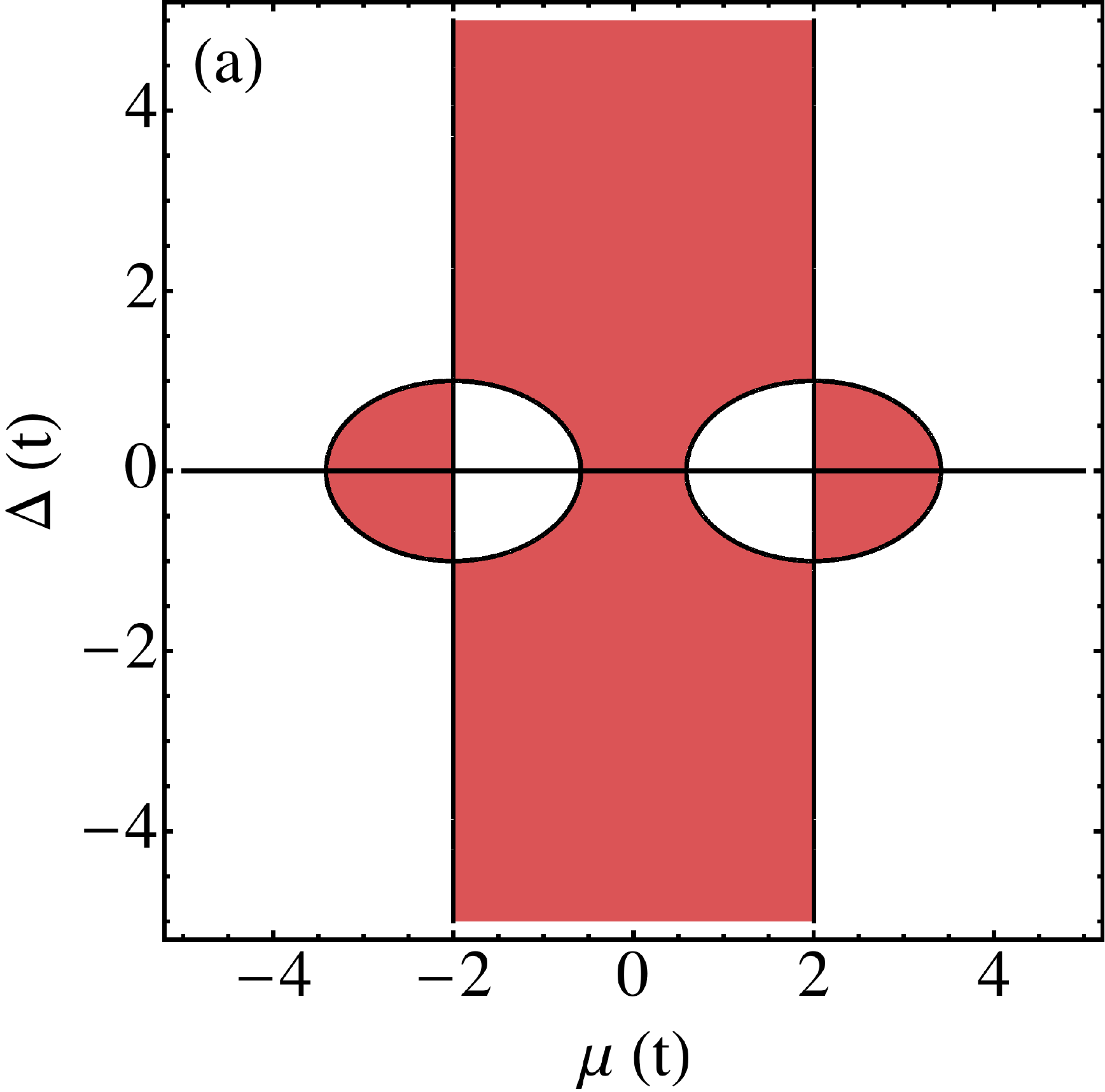}
\includegraphics*[width=0.45\columnwidth]{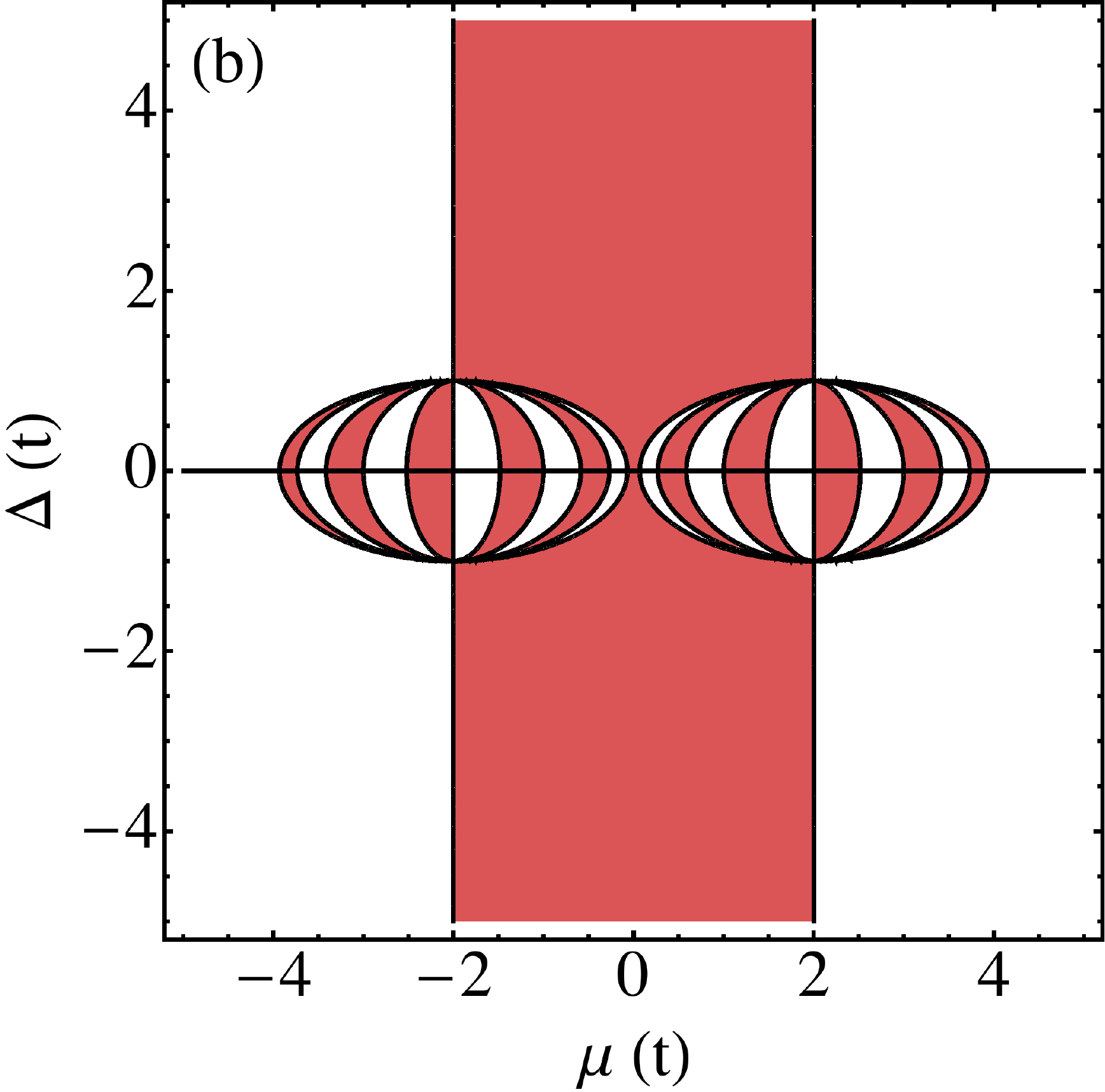}\\
\includegraphics*[width=0.45\columnwidth]{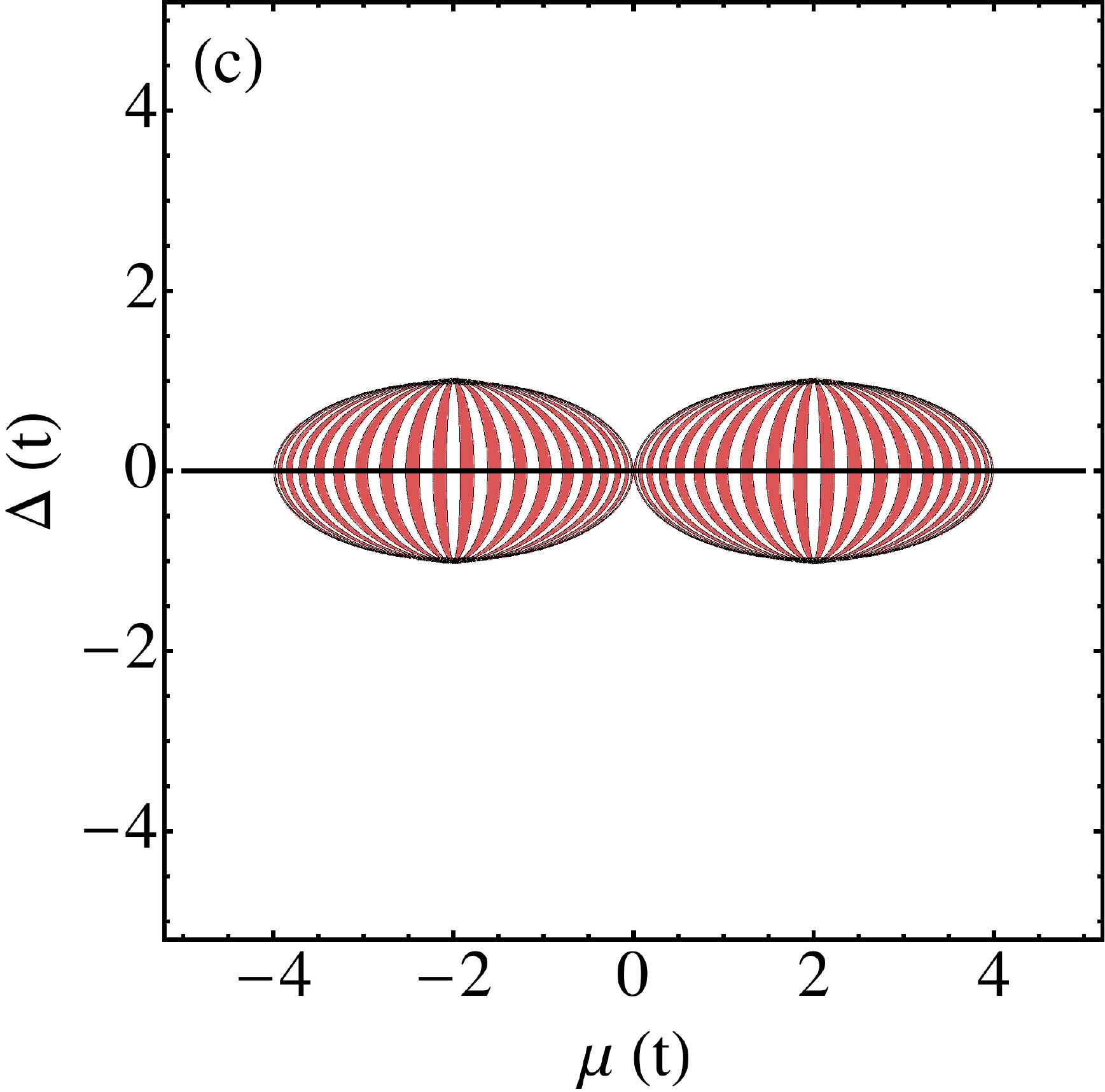}
\includegraphics*[width=0.45\columnwidth]{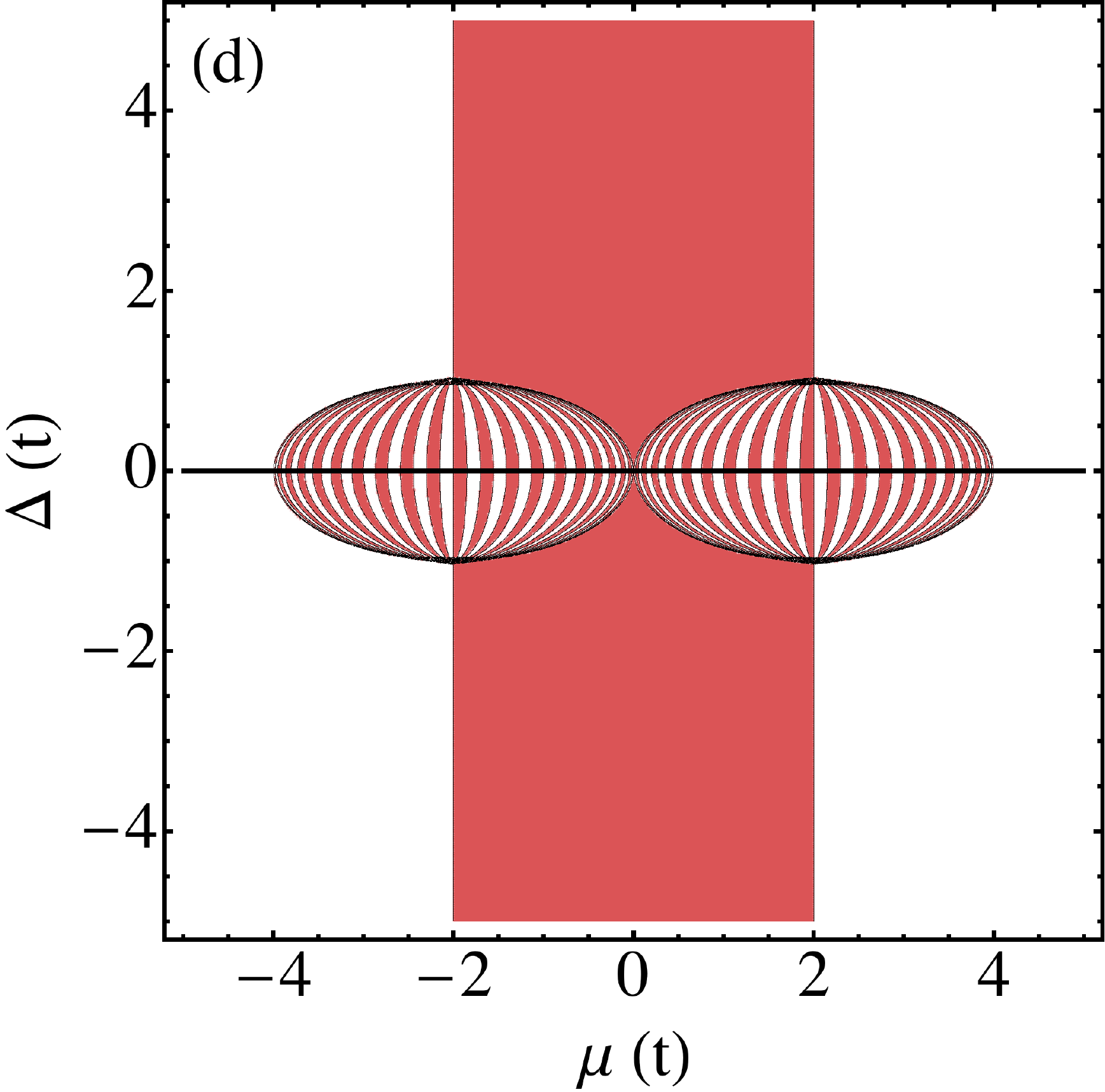}
\caption{(Color online) Topological phase diagram as a function of $\Delta$ and $\mu$ for Kitaev chains with (a) 3, (b) 11, (c) 40, and (d) 41 wires. Light red is the topologically non-trivial phase and white is the topologically trivial phase. The black lines give the points where the bulk gap closes.}
\label{gapclosingladders1}
\end{figure}

We want to study the formation and destruction of the MBS in open quasi-1D and 2D systems described by this model. The quasi-1D systems have open boundary conditions (BC) imposed in the $y$ direction. We know that a purely 1D Kitaev chain is topologically non-trivial for $|\mu|<|2t|$ and $\Delta\neq 0$ \cite{Kitaev2001}. Similarly, a 2D $p+\im p$ Kitaev array is topologically non-trivial for $|\mu|<|4t|$ and $\Delta\neq 0$.
For systems of different numbers of wires we first calculate analytically the bulk topological phase diagrams using a topological invariant\cite{Note1,Sato2009a,Sato2009b,Sato2010,Dutreix2014a,Sedlmayr2015c}. The detailss of this calculation are presented in Appendix A. We should stress that this is the first exact calculation of the phase diagram for the quasi-1D Kitaev chains. The resulting phase diagrams are shown  in Fig.~\ref{gapclosingladders1}. Note the difference between the systems with even and odd number of chains. Note also the formation of striped ellipsoidal regions close to $\Delta=0$ in which the system can go between a topological and non-topological phase for smaller and smaller steps in the variation of the parameters when increasing the number of wires.

\begin{figure}
\includegraphics*[width=0.45\columnwidth]{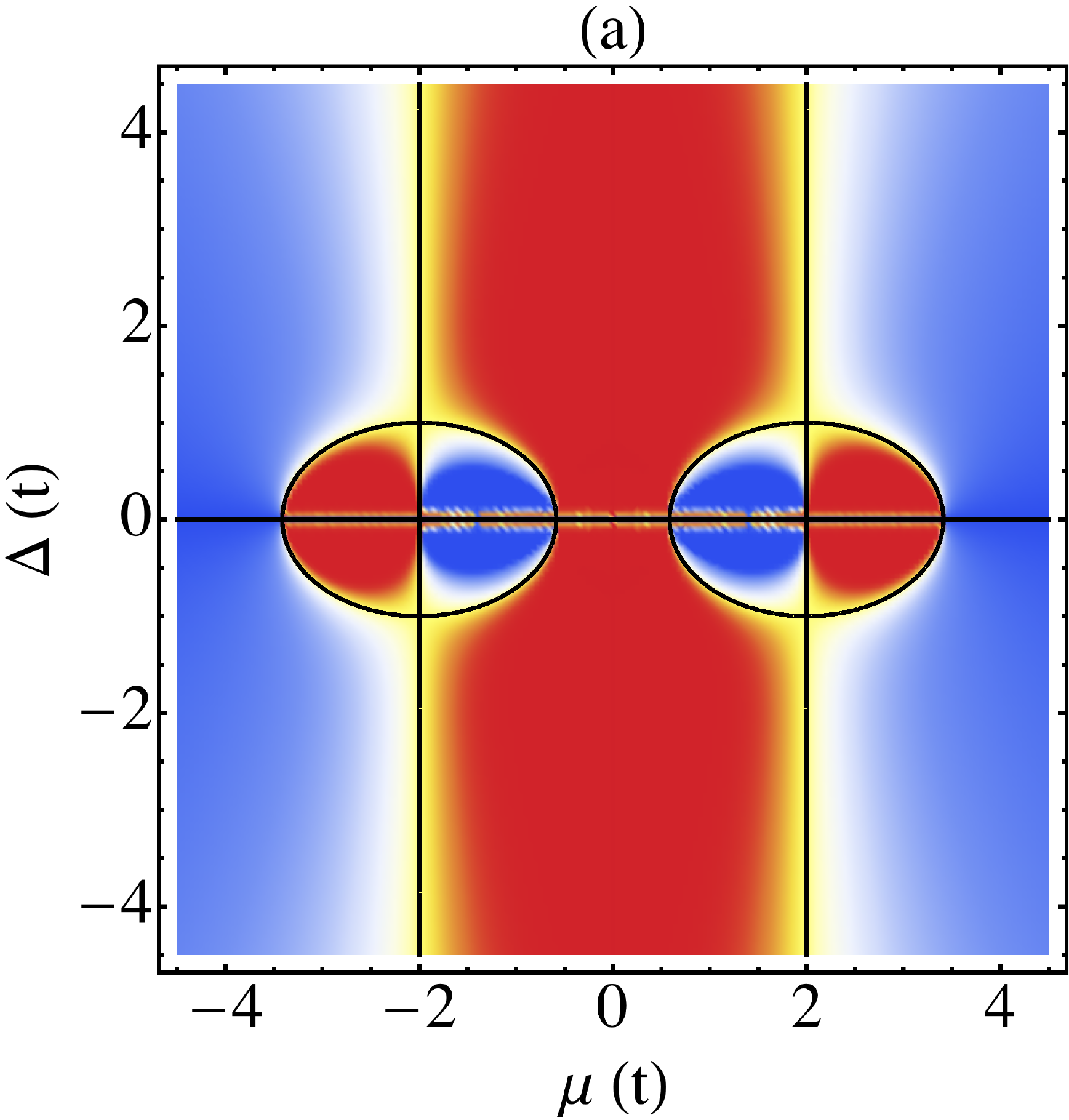}
\includegraphics*[width=0.45\columnwidth]{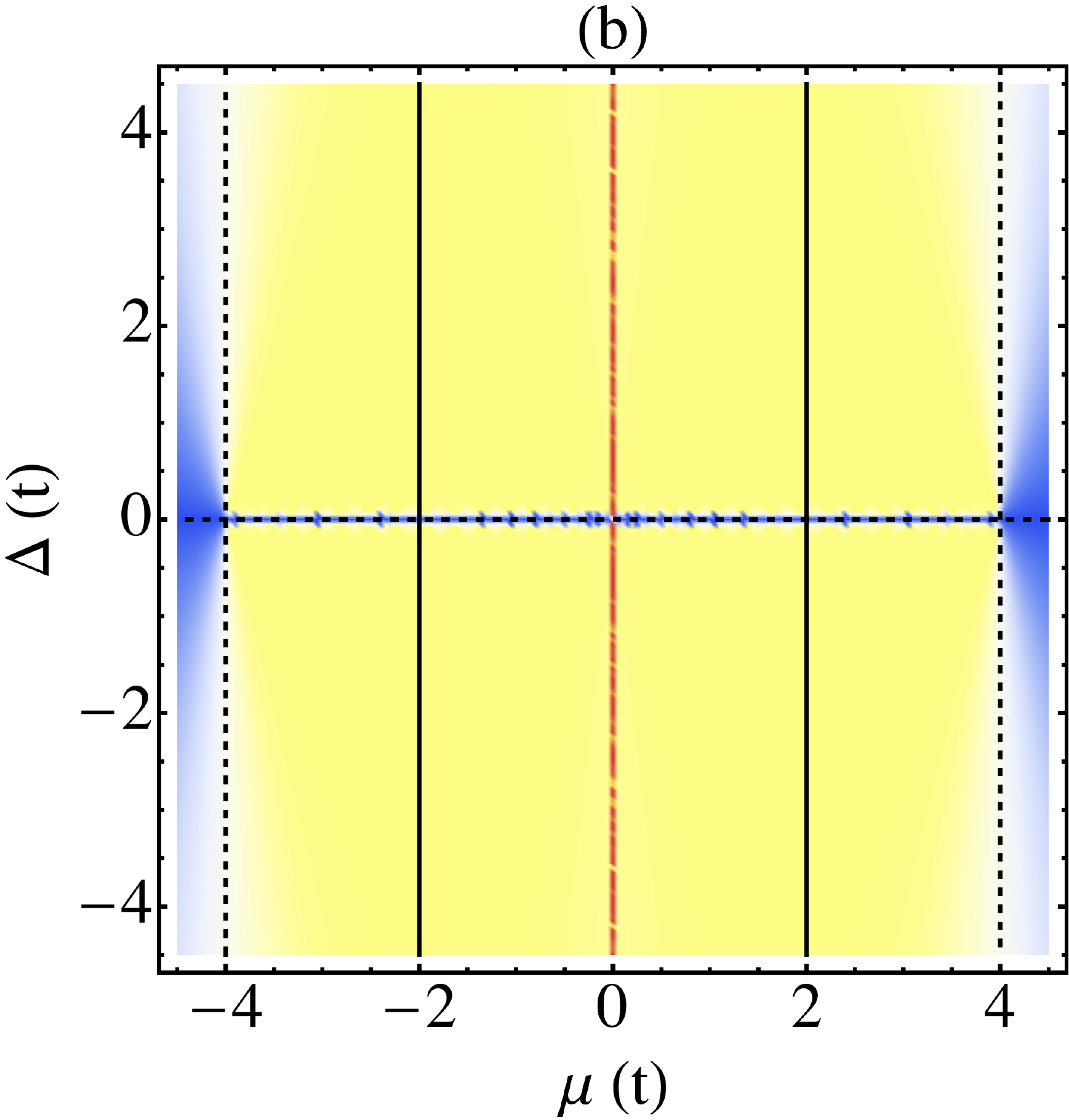}\\
\includegraphics*[width=0.85\columnwidth]{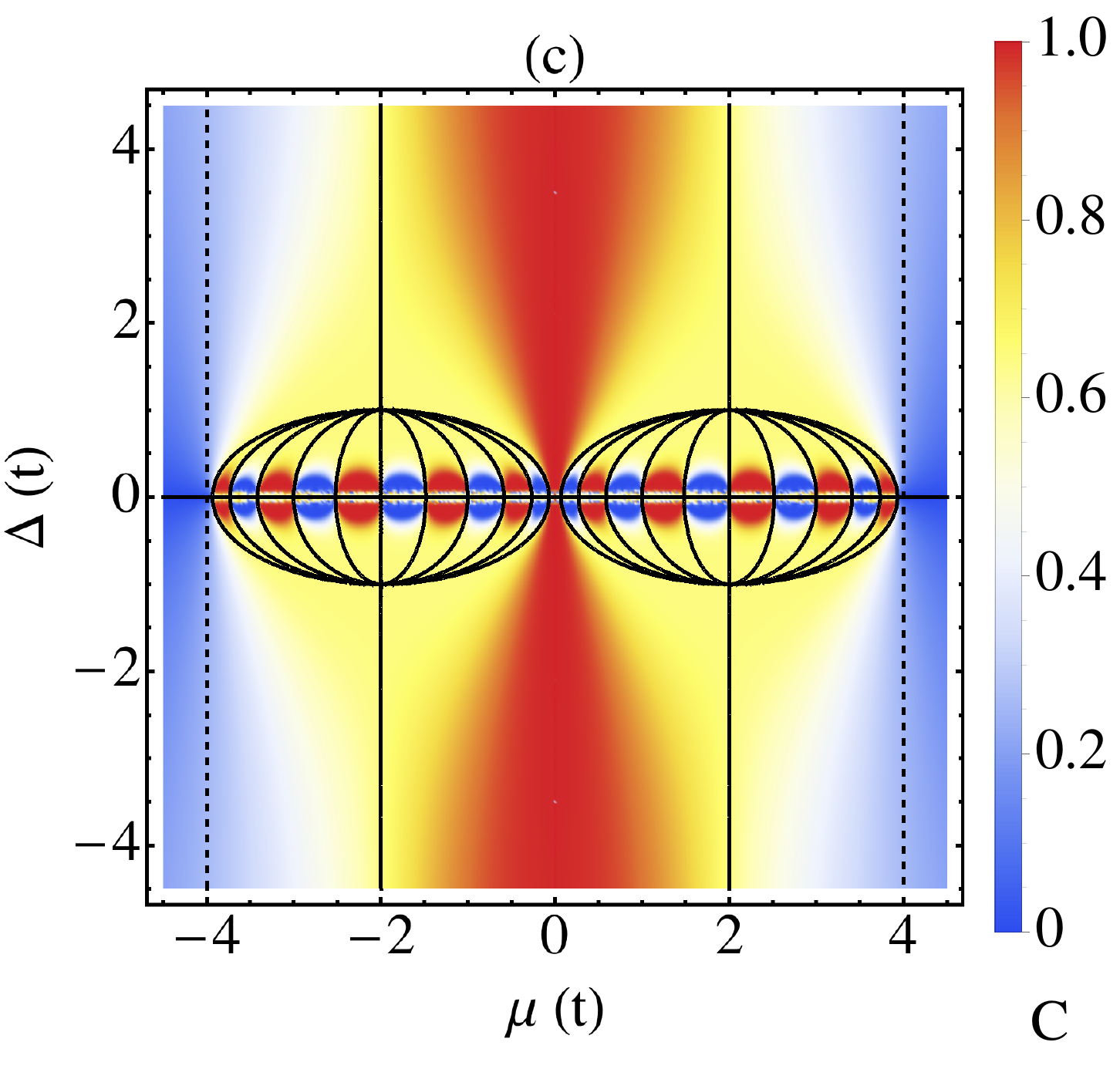}
\caption{(Color online) Total MP (in the left half of the system) for the lowest energy state as a function of $\Delta$ and $\mu$ for Kitaev ladders of size (a) $3\times51$, (b) $41\times41$, and (c) $11\times51$. The solid black lines correspond to the topological phase transitions for the quasi-1D systems, the dashed black lines to the topological phase transitions for the 2D system.}
\label{gapclosingladders2}
\end{figure}

We now compare these analytical phase diagrams with similar ones obtained using a numerical calculation of the MP in finite-size systems (see Fig.~\ref{gapclosingladders2}). We evaluate the total polarization summed over half the system ($\mathcal{R}$) and normalized by the total DOS, see Eq.~\eqref{mp}, which should be equal to $1$ in the case of a MBS. The local components of the wave function $u_j$ and $v_j$ on a site $j$ are obtained by performing a numerical exact diagonalization of the Kitaev Hamiltonian in Eq.~\eqref{kit_ham}, for a finite-size system with open BCs. We can see clearly that there are regions in the phase diagram in which a total MP of $1$ is achieved (denoted in red), which correspond to the regions predicted by the bulk topological phase diagram, see for example Fig.~\ref{gapclosingladders2}(a) for a $3\times51$ site ``ladder'', corresponding to the formation of Majorana states at the ends of the wire. 

\begin{figure}
\includegraphics*[width=0.95\columnwidth]{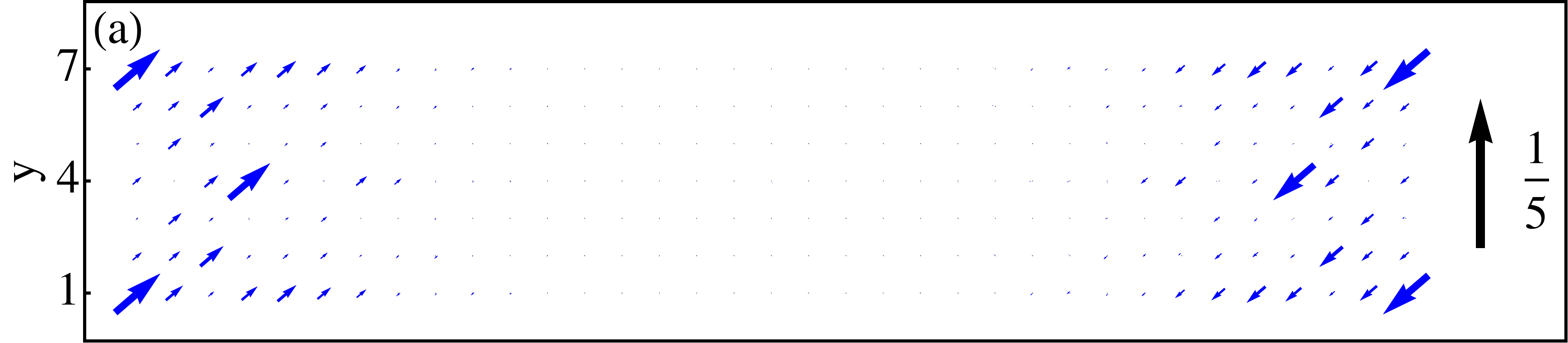}\\
\includegraphics*[width=0.95\columnwidth]{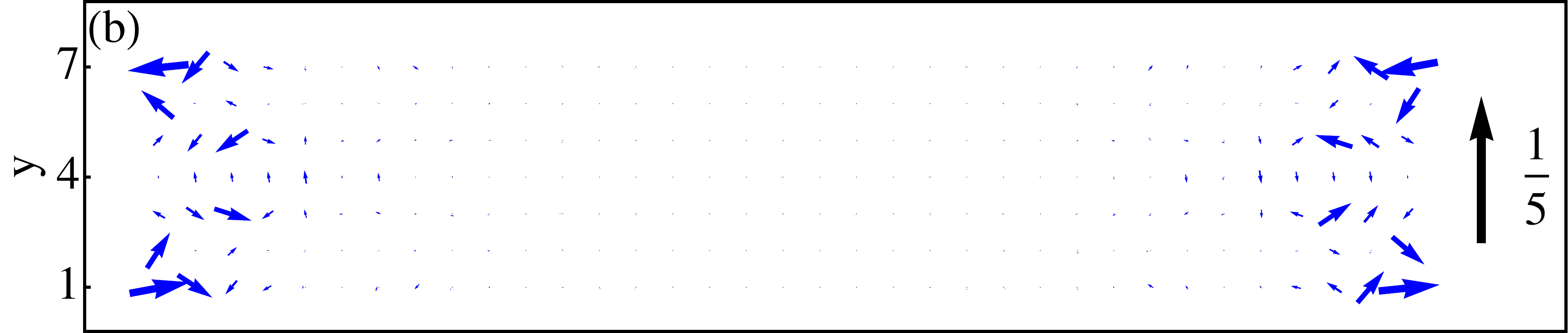}\\
\includegraphics*[width=0.95\columnwidth]{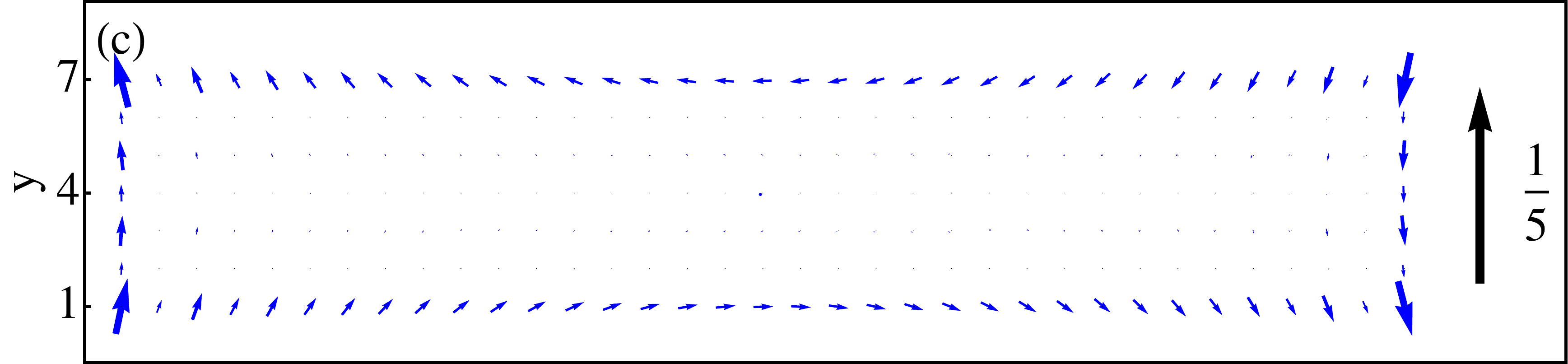}\\
\includegraphics*[width=0.95\columnwidth]{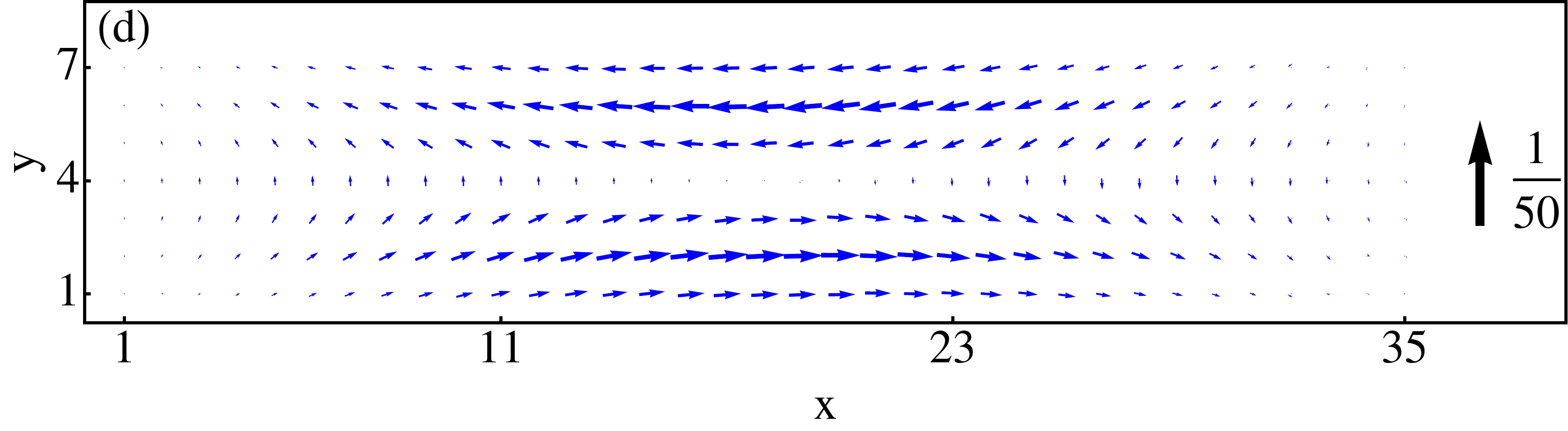}
\caption{(Color online) The MP as a function of position for a $7\times 35$ open system: (a) a MBS (red) for $\mu=1.15t$, $\Delta=0.2t$; (b) a non-MBS (blue) for $\mu=1.65t$, $\Delta=0.15t$; (c) an intermediate state (yellow) $\mu=1.8t$, $\Delta=2t$; (d) a bulk state for a topologically trivial system (blue) for $\mu=4.2t$, $\Delta=0.5t$. The length of the arrows is proportional to the MP, with a scale given by the (black) arrows on the right hand side.}
\label{MBS_Phases}
\end{figure}

For wider ladders however, the formation of topologically protected states along the lateral edges of the ladders, as predicted from the 2D bulk phase diagram, makes things more complicated. These bands tend to vastly reduce the gap and make the interpretation of the phase diagrams tricky, and one may need to consider very long ladders to see a behavior identical with the analytical phase diagrams. For shorter wires, as is the case in Fig.~\ref{gapclosingladders2} this gives rise to the extra yellow regions in the phase diagrams, corresponding to a MP of $0.6-0.8$.
In the Kitaev ladders these intermediate regions arise in substantially large parameter regimes even for a small number of coupled wires. To illustrate this, in Fig.~\ref{MBS_Phases} we present the structure of different states for a $7\times35$ lattice, in different regions of the phase diagram: (a) a purely Majorana state (red, for $\mu=1.15t$, $\Delta=0.2t$), (b) a non Majorana state (blue, for $\mu=1.65t$, $\Delta=0.15t$) (c) an intermediate edge state (yellow, for $\mu=1.8t$, $\Delta=2t$), and (d) a fully non-topological state (blue, outside the 2D topologically non-trivial phase $\mu=4.2t$, $\Delta=0.5t$). Plotted is the complex local Majorana polarization as a 2D vector. Note that in the `pure' Majorana state the MP vector is fully aligned (`ferromagnetic'), while in the fully non-topological state it is locally very small, delocalized in the bulk and disordered. For a `blue' state inside the topological phase, the MP is locally large, but it sums up to zero. In the intermediate (`yellow') states it is localized on the edges and shows locally a full Majorana character, but its direction varies from site to site, making the sum of the MP non-zero but finite.

\begin{figure}
\includegraphics*[width=0.48\columnwidth]{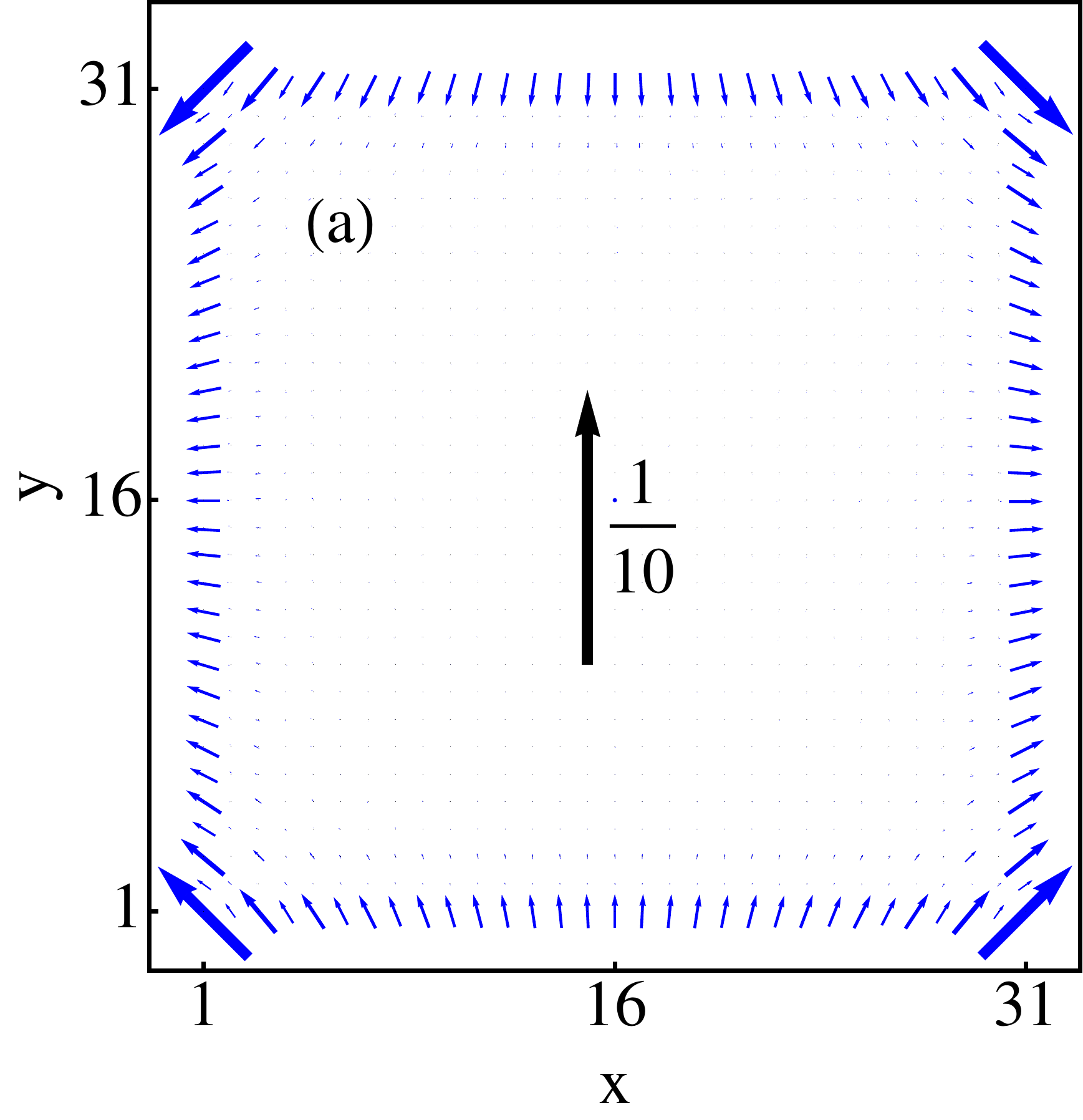}
\includegraphics*[width=0.48\columnwidth]{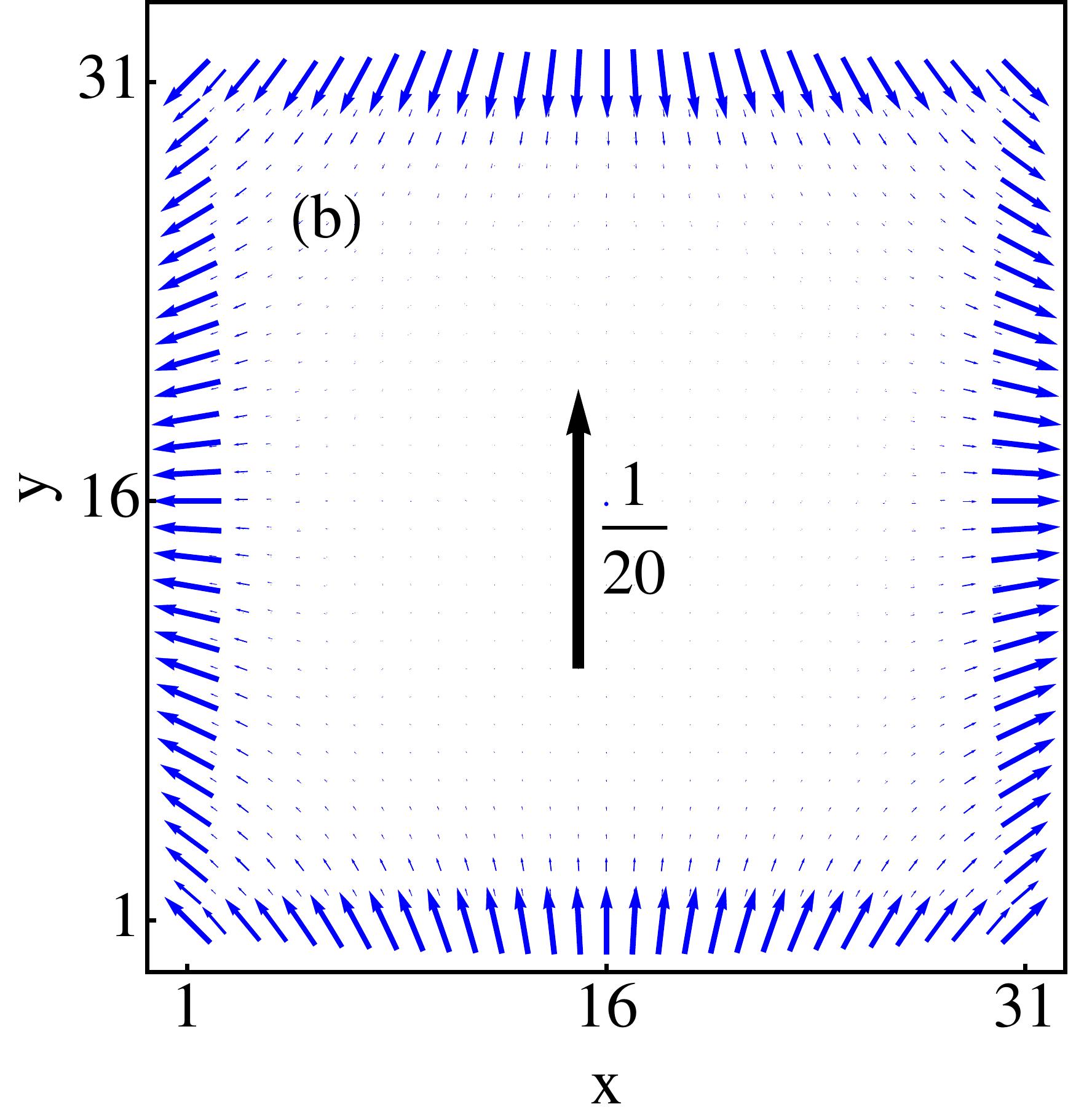}
\caption{(Color online) The MP as a function of position for a $31\times 31$ open system for $\Delta=2t$ and (a) $\mu=1.3t$ and (b)  $\mu=3t$.}
\label{MBS_Phases2}
\end{figure}

When approaching the square system we see that the MP is correctly recovering the boundaries of the phase diagrams for bulk 2D systems, with a value close to $1/\sqrt{2}$. This value can be easily understood by noting that most of the contribution comes from two corner MP vectors of magnitude $1/2$ each, perpendicular to each other (see Fig.~\ref{MBS_Phases2}). Inside the 1d phase transition lines, denoted by the full lines, the topology of the states is described in Fig.~\ref{MBS_Phases2}(a) (the MP is localized at the corners), while in the region between the 1D and 2D boundaries by the structure in Fig.~\ref{MBS_Phases2}(b) (the MP is localized along the edge of the system). This shows that the MP can capture the topology of the 2D Majorana-like states, as well as their origin, be it 1D or 2D bulk topology.

The formation of non-zero energy states which are locally Majorana but for which the direction of the MP vector is varying spatially is in fact a generic feature for all high energy states in `topological bands', for example the bands of Andreev bound states in topological SN junctions (see the SI for a description of such states in terms of the MP), or the bands of edge states in topological 2D systems (see e.g.~Ref.~\onlinecite{Dutreix2014} and references therein). Such bands, for which the lowest energy state is a Majorana, can be thought of as topological in character, since the higher energy states also show some large local Majorana character, which is however not `ferromagnetic', but varies spatially. 
Here we have for the first time an appropriate tool to understand the topology and the structure of such states. It would be interesting in the future to study the properties and the usefulness for applications of such quasi-Majorana states.

\section{Spin-full models}\label{sec_spin_def}

Let us now consider a spin-full state written in the Nambu basis:  $\Psi^\dagger_j=\{c^\dagger_{j\uparrow},c^\dagger_{j\downarrow},c_{j\downarrow},-c_{j\uparrow}\}$, where $c_{j\sigma}^{(\dagger)}$ annihilates (creates) a particle of spin $\sigma$ at site $j$. The corresponding wavefunction is $\psi^T_i$: $\{u_{j\uparrow},u_{j\downarrow},v_{j\downarrow},v_{j\uparrow}\}$. The particle hole operator is $\C=\e^{\im\zeta}{\bm \sigma}^y{\bm \tau}^y\hat K$, where $\hat K$ is the complex-conjugation operator, and $\zeta$ is an arbitrary phase. We will use $\vec{\bm\tau}$ to denote the Pauli matrices in the particle-hole subspace and $\vec{\bm\sigma}$ as the Pauli matrices in the spin subspace. A MBS, $\gamma$, is by definition a state which satisfies $\C\gamma=\e^{\im\tilde\zeta}\gamma$ with $\tilde\zeta$ an arbitrary phase.

Irrespective of which spin basis we choose, our test becomes $v^*_{j\sigma}=-\sigma \e^{2\im\phi_{j\sigma}}u_{j\sigma}$ and $\phi_{j\sigma}=\phi$ must be both spatially and spin independent in the region where the Majorana is localized. Exactly as for the spinless case this arbitrary phase, which cannot be physically fixed, does not affect the properties of the Majorana state and we can choose it in a convenient manner.

As before we can consider the local MP vector
\begin{equation}
\langle\Psi|\C_j|\Psi\rangle=-2\sum_\sigma \sigma u_{j\sigma}v_{j\sigma}
\end{equation}
Note that the condition to have a Majorana state is unchanged from the spinless case and is given in Eq.~(\ref{mp})

\subsection{Generic spin-full model}

We consider the one dimensional tight-binding Hamiltonian for a chain of $N$ sites
\begin{eqnarray}\label{hamiltonian}
H&=&-\frac{1}{2}\sum_{x=1}^{N-1}\Psi^\dagger_x\left[t_x+\im\alpha{\bm \sigma}^y\right]{\bm \tau}^z\Psi_{x+1}+\textrm{H.c.}
\\&&+\sum_{x=1}^N\Psi^\dagger_x\left[-(\mu-t){\bm\tau}^z-\Delta{\bm\tau}^x+B\hat{n}_x\cdot\vec{\sigma}_{\sigma\sigma'}\right]\Psi_{x}\,,\nonumber
\end{eqnarray}
$t_x$ is the nearest neighbor hopping strength which is allowed to vary spatially, $\mu$ is the chemical potential, $B$ is an applied Zeeman field, $\Delta$ is the s-wave superconducting pairing assumed to be induced by a proximity effect, and $\alpha$ is the Rashba spin-orbit coupling. The magnetic field direction is allowed to vary as a function of position:
\begin{equation}\label{mag}
\hat{n}_x=(\cos\vartheta_x\sin\varphi_x,\sin\vartheta_x\sin\varphi_x,\cos\varphi_x)\,.
\end{equation}

\begin{figure}
\includegraphics*[width=0.95\columnwidth]{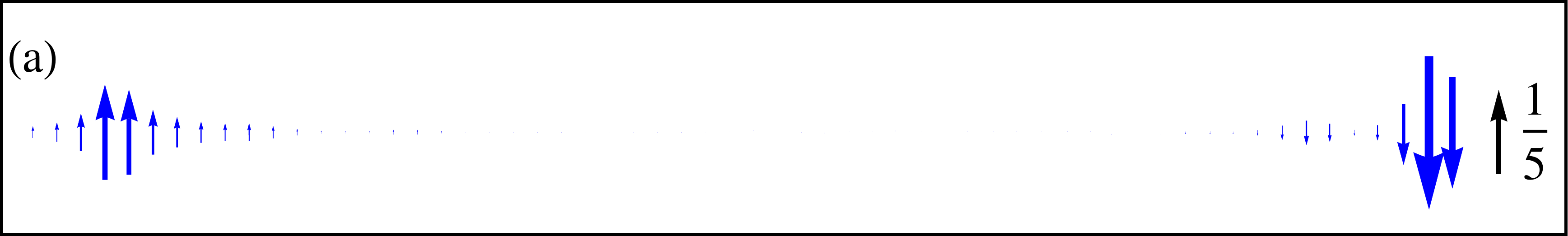}\\
\includegraphics*[width=0.95\columnwidth]{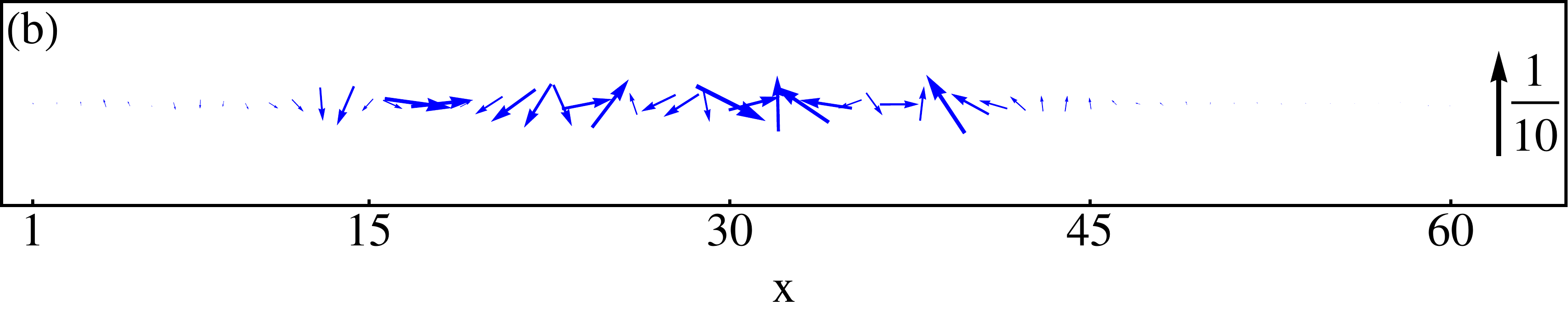}
\caption{(Color online) The MP as a function of position for a disordered system with $\varphi_x=0.3\pi(j-1)$, $\vartheta_x=0.1\pi(j-1)$, $t_x=t+0.2t\tanh[(i-1)/N]$, $\mu=0$, $N=60$, $\Delta=0.3t$, $\alpha=0.01t$, $B=0.4t$. Panel (a) shows the lowest energy Majorana state and (b) the second energy state.}
\label{PH_Mess}
\end{figure}

To exemplify the stability of the generalized MP we focus on a very complicated system for which $\varphi_x=0.3\pi(j-1)$, $\vartheta_x=0.1\pi(j-1)$, $t_x=t+0.2t\tanh[(i-1)/N]$, $\mu=0$, $N=60$, $\Delta=0.3t$, $\alpha=0.01t$, and $B=0.4t$. We also add a specific realization of disorder to both the onsite chemical potential and to the hopping $t_x$\footnote{A specific disorder realization is implemented using random number generators. For the system of $N=60$ sites we add to 12 randomly assigned sites an onsite electronic potential which varies randomly between $0\to0.4t$. Similarly, on 30 randomly assigned nearest neighbor bonds we modify the hopping by an amount which varies randomly between $0\to0.3t$.}. By exactly diagonalizing our system we find the eigenvalues and the eigenstates and we test that we have indeed a Majorana fermion forming. Thus in Fig.~\ref{PH_Mess}(a) we plot the MP for the lowest energy state, which is very close to zero, as a function of position; indeed we observe the ordered `ferromagnetic' Majorana states forming in each half of the wire. We have checked that, remarkably enough, this state satisfies Eq.~\eqref{mp} even if many symmetries of the problem are broken. In Fig.~\ref{PH_Mess}(b) we plot the MP for the state corresponding to the second energy level, and we see that the Majorana vector is fully disordered in this case. We do not show it here but we have checked that by plotting separately the MP for each individual spin that the `ferromagnetic' character is preserved, and that the MP does not depend on the spin basis we have chosen.

\subsection{The Majorana polarization applied to SN junctions}

\begin{figure}
\includegraphics*[width=0.95\columnwidth]{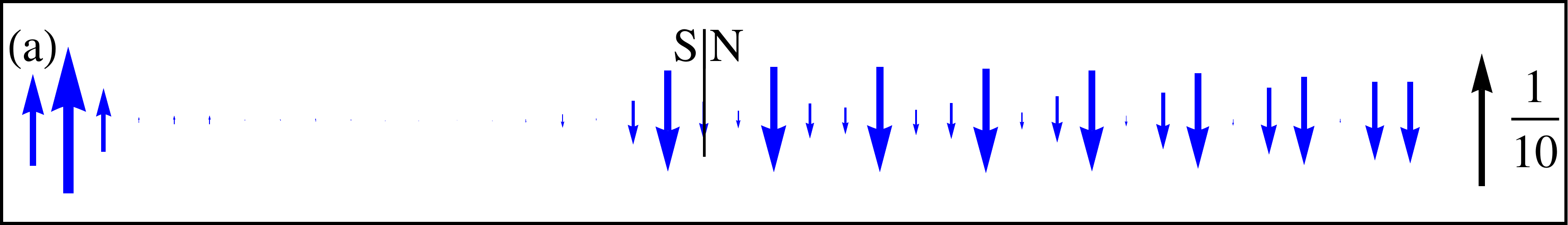}\\
\includegraphics*[width=0.95\columnwidth]{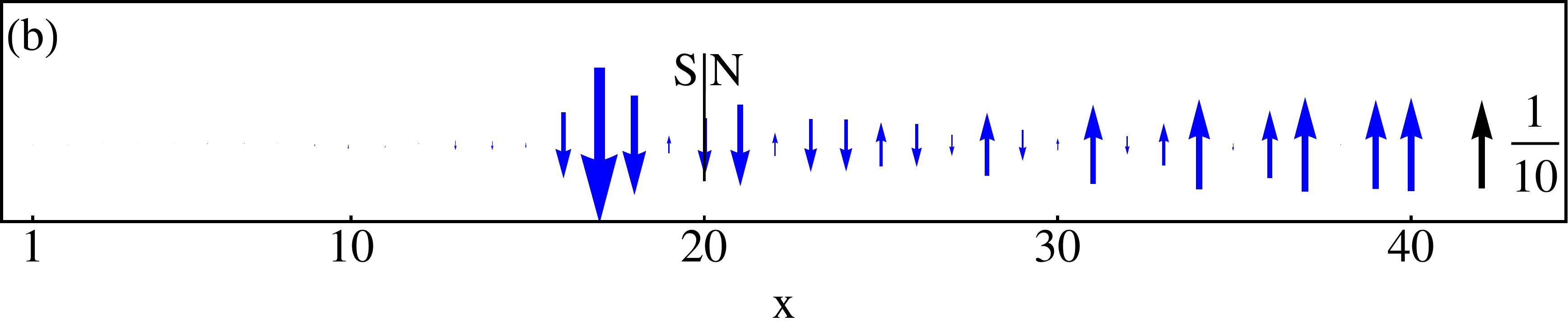}
\caption{(Color online) Local Majorana polarization as a function of position for an SN junction with $\varphi_x=\vartheta_x=0$, $t_x=t=1$, $\mu=0$, $\alpha=0.2t$, $B=0.5t$, $N=40$, and $\Delta_{x>20}=0.4t$ in the S region and $\Delta_{x\leq20}=0$ in the N region. (a) shows the lowest energy Majorana state and (b) the next energy state an ABS. The MBS contribution at the superconductor edge near $x\approx1$, has been scaled down by $1/4$ relative to the extended Majorana in the N region so that they can be shown in the same figure. }
\label{SN}
\end{figure}

We also present the MP for the two lowest energy states in a superconducting-normal (SN) junction: the zero energy Majorana state and the first Andreev bound state (ABS). In Fig.~\ref{SN} we focus on the example $\varphi_x=\vartheta_x=0$, $t_x=t=1$, $\mu=0$, $\alpha=0.2t$, $B=0.5t$, $N=40$, and $\Delta_{x>20}=0.4t$ in the S region and $\Delta_{x\leq20}=0$ in the N region. We see that the lowest energy state exhibits a localized Majorana in the SC and an extended uniform Majorana in the normal state, as predicted in \cite{Sticlet2012}. More interestingly, the first ABS, while showing locally a large MP, is not a Majorana, as this polarization oscillates along the wire and we cannot find any region $\mathcal{R}$ over which its integral can be equal to its integrated density. The next higher energy states all show a similar behavior, with increasing numbers of nodes in the MP oscillations.

\section{Relation of generalized MP to the original MP definition}
\label{oldmpdef}
\begin{figure}
\includegraphics[width=0.85\columnwidth]{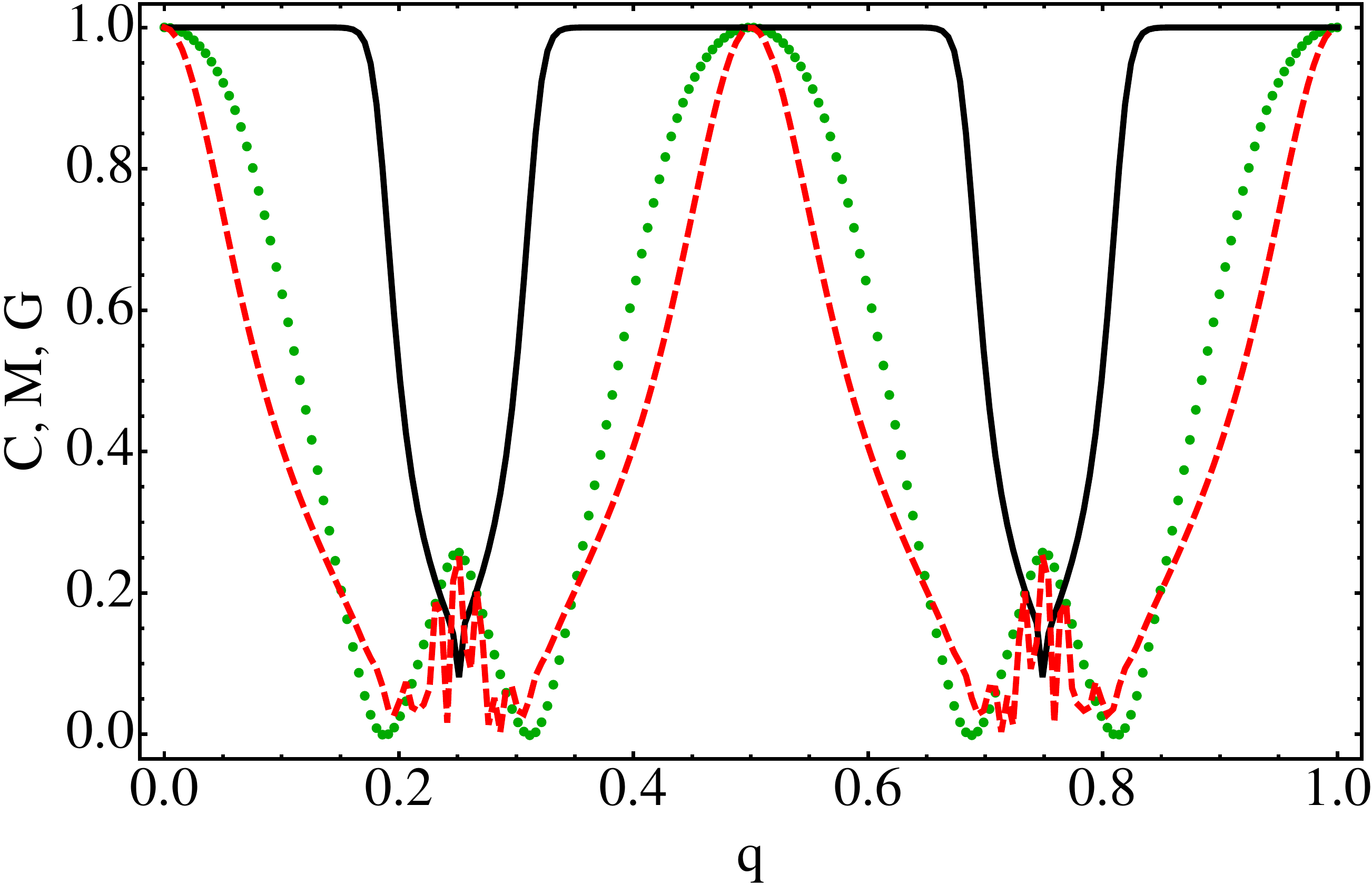}
\caption{(Color online.) Total Majorana polarization of the lowest energy state inside $\mathcal{R}$, the left half of the wire, as a function of the precession $q$. Here $\varphi_x=\pi/2(x-1)$, $\vartheta_x=2\pi q(x-1)$, $t_x=t$, $\mu=3t$, $\alpha=0.2t$, $B=2t$, $N=80$, and $\Delta_{x}=0.5t$. Solid (black) lines show the updated form, $C$, and the dashed (red) lines show the original form $M$. The dotted (green) lines show, $G$, the gap renormalized by the gap at $q=0$.
}
\label{Maj_Pol_Rotating_Field}
\end{figure}

To see the effects of redefining the MP,  in Fig.~\ref{Maj_Pol_Rotating_Field} we show the normalized MP in both original form introduced in Ref.~\onlinecite{Sticlet2012}, and in the corrected form given in the present work, for the lowest energy state as a function of the precession speed $q$ where $\vartheta_x=2\pi q(x-1)$, see Eq.~\eqref{hamiltonian}.
We compare $C$ as defined here to the form of the MP in Ref.~\onlinecite{Sticlet2012} given by
\begin{equation}\label{mpold}
M=\frac{\left|\sum_{j\in \mathcal{R}}\langle\Psi|\M_j|\Psi\rangle\right|}{\sum_{j\in \mathcal{R}}\langle\Psi|{\hat r}_j|\Psi\rangle}\,,
\end{equation}
where\cite{Sticlet2012}
\begin{equation}\label{mpold}
\M_j=\left({\bm\tau}^y{\bm\sigma}^y+\im{\bm\tau}^x{\bm\sigma}^y\right){\hat r}_j\,.
\end{equation}
Although both quantities show a suppression close to the points where the gap closes, the generalized MP form captures correctly the formation of the Majorana states and is equal to $1$ when such states form, in contrast with the original MP form which stays finite but not equal to $1$ except for a few special points. The position of the topological phase transitions given by the generalized MP criterion is in agreement with the points at which the gap closes. To find the bulk gap we use the following heuristic formula:
\begin{equation}
G=\frac{\epsilon_2(q)-\ell(q)}{\epsilon_2(q=0)-\ell(q=0)}\,,
\end{equation}
where $\epsilon_2(q)$ is the second positive energy level and $\ell(q)$ is the mean level spacing.

This system is generically in the D class, however it falls into the BDI class at three points: $q=0,1/2,1$. 
It is precisely at these three points at which the old Majorana polarization can be used. Note that there are other BDI realizations where the original MP formula in Eq.~\eqref{mpold} would however not work.

Note also that for a system with a uniform phase gradient and a total phase difference of $\pi$ \cite{Chevallier2013a} the generalized MP introduced here would correctly recover the formation of two Majorana fermions with opposite polarization, while using the old MP such as in Ref.~\cite{Chevallier2013a} one would obtain two Majorana fermions with the same polarization. The present form is clearly accurate in capturing the conservation of the MP, however, as detailed above, it cannot capture an overall phase factor, so it cannot keep track of the SC phase.

\section{Conclusions}\label{conc}
We present a generalized definition for the Majorana polarization describing locally the Majorana character of a given state. We apply it to a 2D finite-size Kitaev system and to a 1D topological SC wire.  We show that the spatial structure of our local order parameter is a sufficient criterion to distinguish a Majorana state from a non-Majorana state and that the criterion of small energy is not sufficient to prove the Majorana character of a state. For example, even for some infinitesimally small energies, the MP may show spatial oscillations which do not allow one to isolate a spatial region over which the total MP integrates to $1$ (the characteristic of a full Majorana state). The only alternative is a calculation of the bulk invariant or a scaling analysis of the energies with system length, neither of which can be in general easily implemented.

The MP is not directly measurable in any current experiment, since the MP is a Majorana analogue of the particle density of states, and as such the necessary probe would require the injection of an isolated Majorana into the system. However it is an extraordinary versatile and we believe indispensable theoretical tool that allows one to determine the topological character of a given system based on the form of its eigenstates. This is of particular interest for example for determining the topological character of fully open systems for which one does not have other appropriate tools. One interesting observation about such systems which can be obtained solely using the MP, is the existence of non-Majorana topological states which are locally Majorana-like but cannot be integrated to a full Majorana state over a finite region of the system. The formation of these states is generally described by the bulk topological phase diagram of the system. Since the formation of these states depends strongly on the geometry, it would be crucial to investigate their formation in quasi-3D topological wires, to check that, when taking into account realistic parameters, true Majorana states actually can form in InAs and InSb wires. 
It would be also interesting to study the usefulness of such quasi-Majorana or precursor Majorana states for quantum computation, and their braiding characteristics. 

\acknowledgements

This work is supported by the ERC Starting Independent Researcher Grant NANOGRAPHENE 256965. We thank Pascal Simon, Marine Guigou, and Juan Manuel Aguiar for interesting discussions.

\appendix

\section{Topological invariant of quasi-1D Kitaev chain}\label{appa}

We start from a quasi-one-dimensional system with periodic boundary conditions (PBCs) along the bulk $x$ direction and open boundary conditions (OBCs) along the finite $y$ direction, described by Eq.~(2) of the main text. After a Fourier transform along $x$
we can write the Hamiltonian as $H=\sum_{ k}\Psi^\dagger_{ k}\mathcal{H}( k)\Psi_{ k}$ with
\begin{equation}\label{kh}
\mathcal{H}( k)=\begin{pmatrix}
{\bm f}( k) & \CL_{ k}\\
 \CL^\dagger_{ k} & -{\bm f}( k)
\end{pmatrix}\,,
\end{equation}
where
\begin{equation}
{\bm f}( k)=\begin{pmatrix}
f( k) & -t & 0 & 0& \ldots\\
 -t& f( k)  &  -t & 0& \ldots\\
 0 &  -t & f( k)  & -t& \ldots\\
  0 & 0 & -t& f( k) & \ldots\\
  \vdots&  \vdots& \vdots&  \vdots&\ddots
\end{pmatrix}\,,
\end{equation}
and
\begin{equation}
\CL_{ k}=\begin{pmatrix}
\CL_{ k} & \im\Delta & 0 & 0& \ldots\\
-\im\Delta& \CL_{ k} &  \im\Delta & 0& \ldots\\
 0 &  -\im\Delta& \CL_{ k}  & \im\Delta& \ldots\\
  0 & 0 & -\im\Delta& \CL_{ k} & \ldots\\
  \vdots&  \vdots& \vdots&  \vdots&\ddots
\end{pmatrix}\,.
\end{equation}
Finally $f(k) =-2t\cos[k]-\mu$ and $\CL_{ k} =-2\im\Delta\sin[k]$.

In order to calculate the topological invariant we can calculate the parity of the negative energy bands at the time reversal invariant (TRI) momenta, $\Gamma_1=0$ and $\Gamma_2=\pi$. Here the parity refers specifically to a quantity which commutes with the Hamiltonian, but only at the TRI momenta, and anti-commutes with $\C$\cite{Sato2009a}. By transforming to the basis in which the parity operator is
\begin{equation}
P_{N_y}=\begin{pmatrix}
\mathbb{I}_{N_y}&0\\
0&-\mathbb{I}_{N_y}
\end{pmatrix}\,,
\end{equation}
then the Hamiltonian at the TRI momenta is, in this basis,
\begin{equation}\label{blockh}
\tilde{\mathcal{H}}(\hat\Gamma_i)=\begin{pmatrix}
\bar{\mathcal{H}}(\hat\Gamma_i)&0\\
0&-\bar{\mathcal{H}}(\hat\Gamma_i)
\end{pmatrix}\,.
\end{equation}
$\tilde{\mathcal{H}}( k)=\mathcal{U}^\dagger\mathcal{H}( k)\mathcal{U}$ with $\mathcal{U}$ the rotation between Eqs.~\eqref{kh} and \eqref{blockh}.
For the Kitaev chains under consideration the rotation is
\begin{equation}
\mathcal{U}=\frac{1+{\bm\tau}^x}{2}\mathbb{I}_{N_y}+\frac{1-{\bm\tau}^x}{2}\bar{\mathbb{I}}_{N_y}\,,
\end{equation}
where $\bar{\mathbb{I}}_{N_y}$ is the $N_y\times N_y$ matrix given by $[\bar{\mathbb{I}}_{N_y}]_{nn'}=\delta_{n,N_y+1-n'}$.
The topological invariant is
\begin{equation}
\delta=\sgn\det\bar{\mathcal{H}}(\hat\Gamma_1)\det\bar{\mathcal{H}}(\hat\Gamma_2)\,.
\end{equation}
When $\delta=-1$ there is band inversion, i.e.~the parity switches between TRI momenta an odd number of times and the system is topologically non-trivial. For $\delta=1$ the system is topologically trivial. For more information, and how to generalize this calculation to a spinful s-wave chain see Ref.~\onlinecite{Sedlmayr2015c}.

\section{Relation of MP to the chiral Majorana character}\label{app_char}

The local chiral Majorana character, $\V$, introduced in Ref.~\onlinecite{Sedlmayr2015a}, can be written for any BDI system, but not for any other symmetry class. This operator plays for these particular systems a similar role to the particle-hole operator in the present work. However, it has additional properties related to the fact that it cannot be written for a D or DIII topological superconductor (TS). Here we describe the relation between the MP and the previously defined chiral Majorana character.

A spin-full state in the Nambu basis,  $\Psi^\dagger_j=\{c^\dagger_{j\uparrow},c^\dagger_{j\downarrow},c_{j\downarrow},-c_{j\uparrow}\}$, where $c_{j\sigma}^{(\dagger)}$ annihilates (creates) a particle of spin $\sigma$ at site $j$, can be described by the wavefunction $\psi^T_i=\{u_{j\uparrow},u_{j\downarrow},v_{j\downarrow},v_{j\uparrow}\}$.
We will write
\begin{equation}\psi_j=\begin{pmatrix}
|u_{j\uparrow}|\e^{\im\phi_{j\uparrow}-\im\theta_{j\uparrow}}\\
|u_{j\downarrow}|\e^{\im\phi_{j\downarrow}-\im\theta_{j\downarrow}}\\
|v_{j\downarrow}|\e^{\im\phi_{j\downarrow}+\im\theta_{j\downarrow}}\\
|v_{j\uparrow}|\e^{\im\phi_{j\uparrow}+\im\theta_{j\uparrow}}
\end{pmatrix}\,,
\end{equation}
and define $\phi_{j\delta}=\phi_{j\uparrow}-\phi_{j\downarrow}$ and $\theta_{j\delta}=\theta_{j\uparrow}-\theta_{j\downarrow}$.

In the most general case for a spinful BDI model the chiral Majorana character is
\begin{eqnarray}
\V_{j}
&=& 2u_{j\downarrow}  v^*_{j\downarrow}\left[\e^{\im\alpha}\cos^2[\beta/2]+\e^{-\im\alpha}\sin^2[\beta/2]\right]\nonumber\\&&
- 2u_{j\uparrow}  v^*_{j\uparrow}\left[\e^{-\im\alpha}\cos^2[\beta/2]+\e^{\im\alpha}\sin^2[\beta/2]\right]\\\nonumber&&
+2\left( u_{j\uparrow}  v^*_{j\downarrow}- u_{j\downarrow}  v^*_{j\uparrow}\right)\im\sin[\alpha]\sin[\beta]\,.
\end{eqnarray}
The angles $\alpha$ and $\beta$ can be calculated from any two, non-parallel, spin vectors at different spatial points in the system. Writing these as $\vec S_{1,2}$, then the angles are defined by\cite{Sedlmayr2015a}
 \begin{equation}\label{ns}
(\sin\alpha\cos\beta,\cos\alpha,\sin\alpha\sin\beta)=\frac{\vec S_1\times \vec S_2}{|\vec S_1\times \vec S_2|}\,.
\end{equation}
As already noted the Majorana states localized at the boundaries of the system have either $\V_j>0$ or $\V_j<0$ and these are well separated, much like the MP used here. This can be explicitly checked by plotting $\V_j$ as a function of position for the eigenstates.
An arbitrary but homogeneous superconducting phase of $\kappa$ requires the transformation $\V\to \V\e^{\im\kappa}$.

We know that for any Majorana $|\langle\gamma|\C|\gamma\rangle|=1$, and equally that
\begin{equation}
\sum_j\left|\langle\gamma|\V_{j}|\gamma\rangle\right|=1\,.
\end{equation}
It is then a simple task to construct a unitary operator which has this property also for a D or DIII TS by allowing ourselves to locally correct for the operator using a spin rotation. The above spin-rotation can be understood as rotating the system to a reference frame in which $S^y_j=0$, for a {\it particular} state. By locally implementing such a rotation, it can be seen that any zero energy state will satisfy
\begin{equation}\label{maj_density1}
\sum_j\left| \langle\Psi|\tilde\V_j|\Psi\rangle\right|=1\,,
\end{equation}
where
\begin{eqnarray}\label{maj_density2}
 \langle\Psi|\tilde\V_j|\Psi\rangle&=&2( u_{j\downarrow} v^*_{j\downarrow}\e^{\im\alpha_j} - u_{j\uparrow} v^*_{j\uparrow}\e^{-\im\alpha_j})\nonumber\\&=&
2\e^{-\im(\theta_{j\uparrow}+\theta_{j\downarrow})}( |u_{j\downarrow} v_{j\downarrow}|\e^{\im\theta_{j\delta}}\e^{\im\alpha_j}\nonumber\\&&\qquad - |u_{j\uparrow} v_{j\uparrow}|\e^{-\im\theta_{j\delta}}\e^{-\im\alpha_j})\,,
\end{eqnarray}
and
\begin{equation}\label{globaltheta}
\alpha_j=-\tan^{-1}\left[\frac{S^y_j}{S^x_j}\right]=\phi_{j\delta}-\theta_{j\delta}\,.
\end{equation}
Here $S^{j,y,z}_j$ are the local particle spin expectation values of the state $|\Psi\rangle$. Naturally such a definition no longer allows one to make any comparison across states or space. For a BDI system this transformation can be performed globally in an appropriately chosen spin basis.
Eq.~\eqref{maj_density1} is referred to as the chiral density. Note however that although all Majorana states will have a chiral Majorana density of one, not all states with a chiral density close to one must be Majorana states, or even low energy localized states. The PH operator allows us to do better at identifying the Majorana states by also defining the appropriate local phase for a state in a way which is independent of the spin basis used:
\begin{equation}
 \langle\Psi|\C_j|\Psi\rangle=2\e^{-\im(\phi_{j\uparrow}+\phi_{j\downarrow})}( |u_{j\downarrow} v_{j\downarrow}|\e^{\im\phi_{j\delta}}
- |u_{j\uparrow} v_{j\uparrow}|\e^{-\im\phi_{j\delta}})\,.
\end{equation}
This local phase is crucial to the arguments and calculations used in the main text.

\end{document}